# Enhancing Energy Sector Resilience: Integrating Security by Design Principles*

Dov Shirtz[1], Inna Koberman[1,2], Aviad Elyashar[2,3], Rami Puzis[1,2] and Yuval Elovici[1,2]

1 Cyber@BGU, Cyber Labs at Ben-Gurion University of the Negev, Beer-Sheva, Israel
2 Department of Software and Information Systems Engineering, Ben-Gurion University of the Negev, Beer-Sheva, Israel
3 Department of Computer Science, Shamoon College of Engineering, Beer-Sheva, Israel

## Abstract

Security by Design (SbD) is a holistic framework for creating and maintaining systems that are inherently resistant to vulnerabilities and hostile incursions. It encompasses not only technical dimensions (e.g., the secure development of industrial control systems, hardware, firmware, software, and communications) but also "soft" factors such as employee security awareness and a leadership's security-minded culture.

Crucially, SbD is a continuous, iterative discipline that begins at the requirements definition and persists through design, implementation, testing, deployment, maintenance, and decommissioning, enabling the early identification and cost-effective remediation of flaws and fostering stakeholder trust.

By embedding security measures (e.g., continuous testing, robust authentication, architectural safeguards, and best-practice coding) at every stage of the product lifecycle, organizations formalize and strengthen their IT and OT defenses, complementing existing policies and standards (e.g., COBIT, Cybersecurity, and QA frameworks) rather than replacing them.

In this document, we articulate detailed SbD requirements tailored to industrial control systems in the energy sector. While adopting these measures and integrating every energy-sector stakeholder, from OEMs (i.e., Hardware (HW) and software (SW) manufacturers) to integrators, may initially seem onerous, strict adherence is essential to cultivating an SbD-driven ecosystem capable of withstanding evolving cyber threats.

* Supported by the U.S.-Israel Energy Centre managed by the Israel-U.S. Binational Industrial Research and Development (BIRD) Foundation.

**Keywords:** Security by design, energy sector, cybersecurity, industrial control systems.

## Contents

# 1. Preface

U.S. Presidential Policy Directive 21 (February 2013) identified the energy sector as a part of the U.S. critical infrastructure. In recent years, industrial control systems (ICSs) in general and those associated with the energy sector have been the target of numerous cyberattacks [Parachute 2022, Kessem 2021, Fox 2021]. It is thus imperative that the energy sector, as a high-profile and targeted sector [Parachute 2022], adopt new and advanced methodologies, frameworks, and technologies to protect itself from cyberattacks and safeguard energy organizations and their customers from cyberattacks. Although the adoption of these methods will require increased investment in cybersecurity, this expense will be offset by the enhanced security provided by these measures, which will increase the reliability and availability of the services provided by the energy sector [Zografopoulos 2021].

The energy sector is not monolithic; it includes entities such as power plants, electricity storage systems, and virtual electricity generators, each with its own operational technology (OT) and information technology (IT) infrastructures. Each entity in the energy sector is considered critical infrastructure and is vulnerable to cyberattacks. For example, the attack on the Colonial Pipeline in May 2021 caused a shutdown by the company, which led to hefty costs. The attack on Ukraine in 2015 cut power to almost a quarter-million people for nearly six hours. These are just examples of how cybercriminals have managed to bypass security controls and exploit vulnerabilities in IT and OT systems within the energy sector in recent years [Beerman 2023, Aljohani 2022].

Although numerous standards, methodologies, frameworks, and best practices (BPs) exist to enhance cybersecurity in OT and IT environments related to the energy sector, we have identified a need for a comprehensive framework for their implementation within the context of SbD, which is currently lacking.

We address this gap by presenting an SbD framework that establishes the requirements for SbD in the OT and IT infrastructures. We do not aim to replace cybersecurity standards and BPs, but rather to provide a framework for their implementation in the context of SbD. They will serve as the foundation of the framework, guiding the application of the SbD approach throughout the design and product lifecycle.

The proposed SbD framework will contribute to the development of products and systems with enhanced cybersecurity resilience, thereby increasing the ability of organizations in the energy sector to withstand and mitigate cyberattacks, thus improving their safety, reliability, availability, and users' trust.

This paradigm shift involves a shift from focusing solely on availability and safety to a more holistic approach, in which cybersecurity, safety, availability, and integrity requirements are all equally important. This shift arises because digitization introduces cybersecurity issues. Consequently, a lack of cybersecurity measures can compromise safety, availability, and integrity, resulting in unsafe systems, equipment failure, and unintended consequences. For example, a cyberattack that manipulates the ICS sensor readings, causing the system to make incorrect decisions [Stouffer 2023, ISO/IEC 62443, Humayed 2017, Karnouskos 2011, Hollerer 2023, Kornecki 2010].

The suggested paradigm is to consider and implement all security and integrity aspects throughout the life cycle of any HW, SW, and system (i.e., requirement setting, design, coding, testing, manufacturing, integration of the products in the production environment, maintenance, and disposal).

---

After the introductory part (sections 1.1 and 2), which outlines some general reasons why organizations should implement SBD, the rest of this document is divided into two parts. The first part discusses the soft (non-technological) aspects, while the second part addresses the technological (tactical) aspects of the SbD framework. The first part details the soft aspects of SbD, such as the responsibilities of all individuals within an organization, from top management and the board of directors to every employee. It also involves all business partners, distributors, manufacturers, and integrators of HW, SW, and systems in the energy sector.

Consequently, organizations in the energy sector should develop an ecosystem that includes all energy companies, manufacturers of the tools used by these companies, integrators, and business partners.

The second part addresses the tactical aspects related to the technical requirements of the SbD framework, applying to all ecosystem participants. It provides a comprehensive description of the requirements for designing and creating SbD infrastructure. This part lays a foundation for implementing modern cybersecurity mechanisms such as moving target defense (MTD), zero trust (ZT), and blockchain. The requirements are outlined using a general HW, SW, and systems lifecycle.

It should be noted that:

1. The SbD framework is not intended to replace international standards, frameworks, and BPs, but rather to build upon them.
2. The physical premises of organizations in the energy sector are assumed to be extremely secure (see Section 4.4.7).
3. The SbD framework is not static; it is dynamic and continuously evolving in response to changes and advancements in technology, as well as new threats.

## 1.1 The Purdue Model

This document relies on the Purdue reference model [Williams 1994]. The basic Purdue model divides an ICS into six levels (Fig. 01).

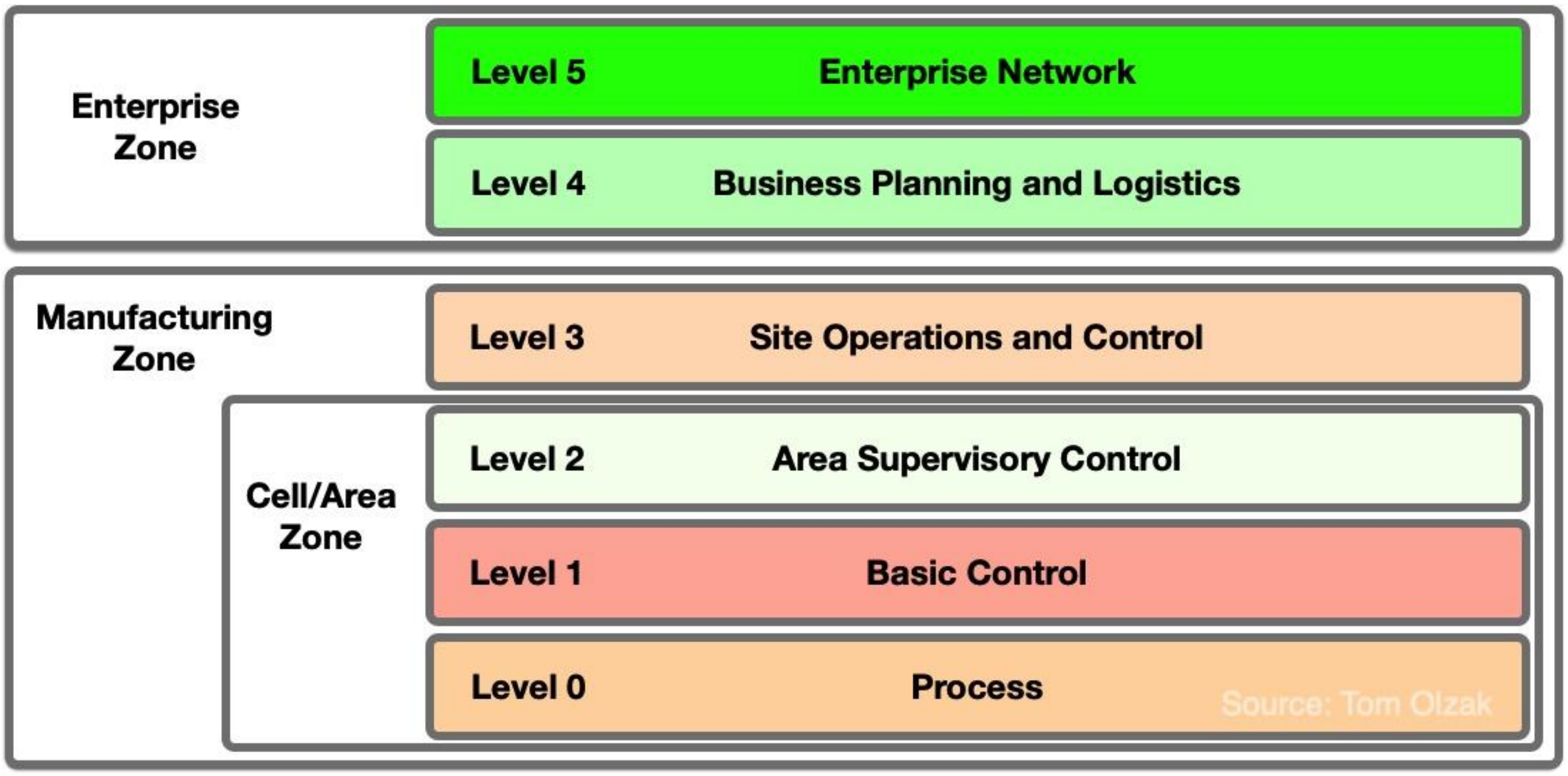

Figure 01: Purdue model schema (from [Olzak 2021]).

- Level 0 comprises physical devices, such as analog sensors and digital sensors (i.e., IoT and IIoT-based devices).
- Level 1 contains components that interact with the physical devices located in level 0, such as a programmable logic controller (PLC).
- Devices in level 2 interact with components located in level 1. These interactions are usually automated, but human operators can also interact with devices via human-machine interfaces (HMIs).
- In level 3, the operational area or zone is defined as the area that controls operations in levels 0-2. The systems at this level collect and manage information from across an entire geographical area.

- In level 4, information from previous levels supports business system functionality related to production, e.g., logistics. IT components at this level include servers, applications, and database servers.
- Level 5 encompasses the organization's IT environment, including all its functions, such as CRM and finance applications.

Due to the necessity of integrating cybersecurity into ICSs, several modifications have been made to the basic Purdue model. These models differ from the original by incorporating a demilitarized zone (DMZ) between levels 3 and 4, which facilitates segmentation, as well as adding firewalls and intrusion detection systems (IDS) between levels 2 and 3. Furthermore, they include advanced logging capabilities, among other improvements (Figure 02) [Ankush 2022, Xu 2022].

Note the following:

1. The commercial off-the-shelf (COTS) market offers a range of products used at levels 0-2.
2. Communication between levels 0-3 is based on protocols such as Modbus and Distributed Network Protocol 3 (DNP3), as well as communication technologies such as 5G.
3. A level 0 device can be either analog or digital. All devices at this level use specific communication methods and protocols to interact with devices, components, or systems at higher levels.

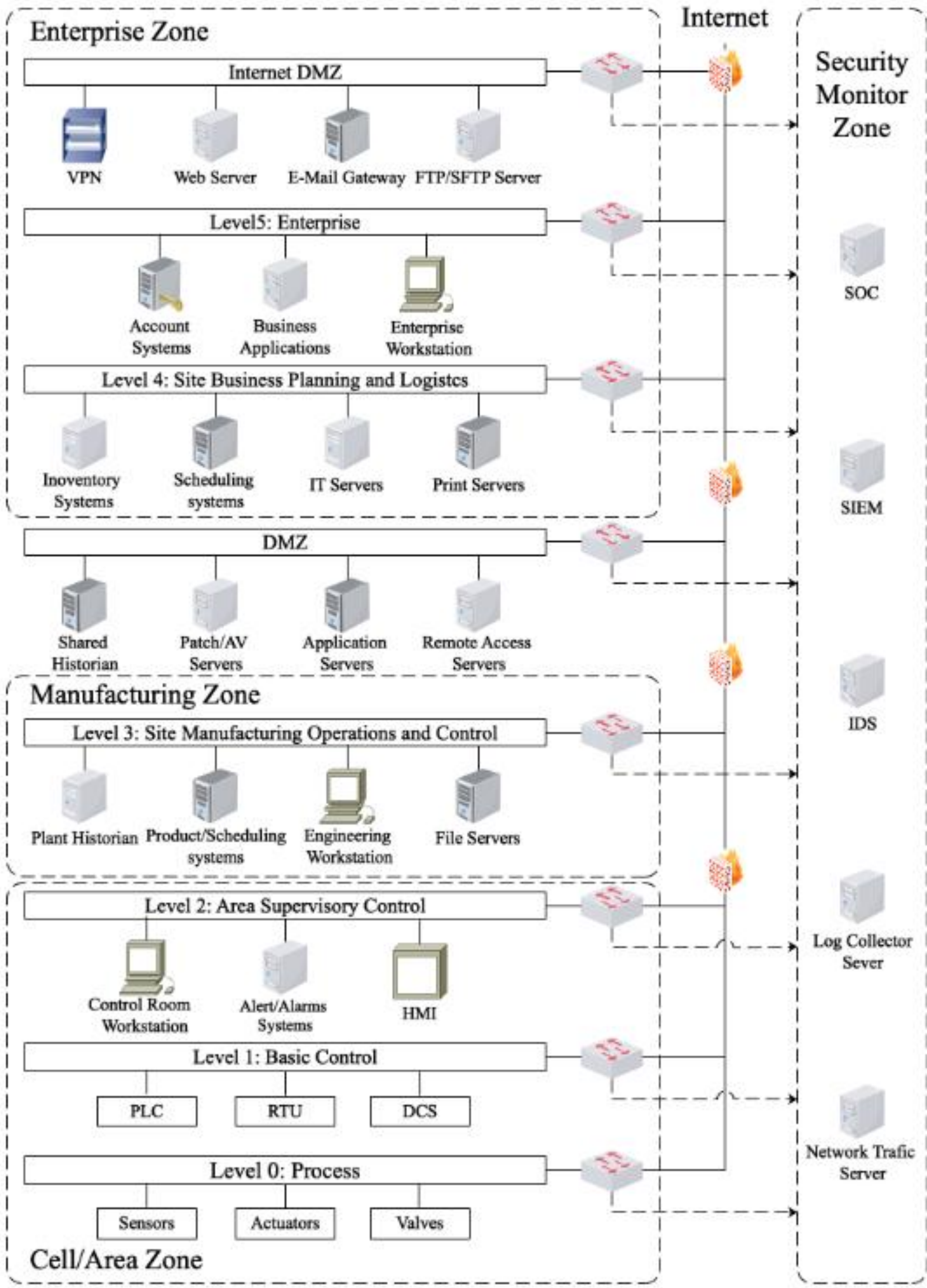

Figure 02. Aspects of cybersecurity are embedded in the basic Purdue model [Xu 2022].

# 2. Preliminaries

## 2.1 Objectives

In this document, we aim to:

1. Establish a framework that enables the energy sector to fulfill the goal of incorporating SbD throughout the ecosystem.
2. Promote compliance among all participants in the energy sector ecosystem with international security standards, frameworks, and best practices.
3. Incorporate international quality assurance (QA) standards, frameworks, and BPs in the SbD framework for all participants in the energy sector ecosystem.

Here are the tasks and responsibilities of the energy sector to achieve the goal of incorporating SbD:

1. Establish a cybersecurity ecosystem (Figure 03) that includes HW and SW manufacturers used by the energy sector, system integrators, business partners, service providers such as Internet providers, as well as all organizations within the energy sector.
2. Establish the following entities, which will be part of the cybersecurity ecosystem:
    a. A computer emergency response team (CERT) and security operations center (SOC). The CERT and SOC should operate at the sector level, i.e., a community CERT, and be connected to all organizations within the ecosystem.
    b. An education, awareness, and research center for cybersecurity. The center will provide high-level cybersecurity education and awareness programs targeted at organizations within the ecosystem, promoting cybersecurity research related to topics of interest in the energy sector. Note that the existence of the center will not replace organizations' responsibility for promoting cybersecurity education and awareness within their organizations.
    c. A central institution for testing, accreditation, and certification, responsible for evaluating products and accrediting companies in the energy sector. This entity will act as a checkpoint, facilitating the entry of a wide range of products

and services into the ecosystem. The existence of this institution does not exempt an organization from conducting on-premises tests.

3. Establish security practices concerning design, development, acquisition, maintenance, and operation of information resources. These resources encompass information from both the OT and IT environments, as well as from internal and external sources.
4. Ensure the integrity of OT and IT products and services by creating a robust infrastructure for testing and validating their security (see 2.c above).
5. Promote continuous cybersecurity risk assessment.
6. Promote the implementation of the proposed SbD framework, particularly in organizations that did not initially build their systems with security at the forefront.

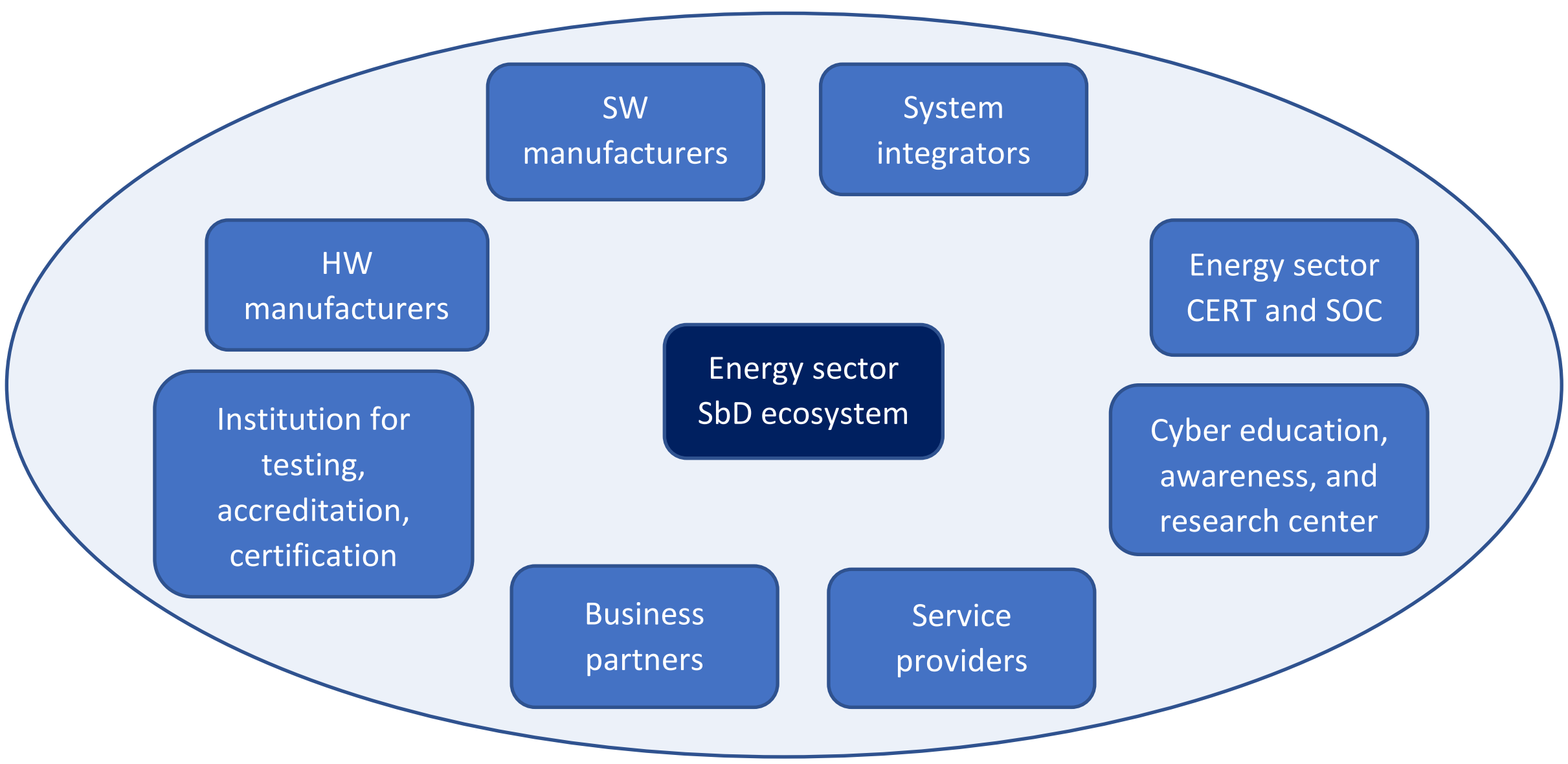

Figure 03: Entities in the energy sector SbD ecosystem.

One might ask what incentives an organization has to invest in, join, and cooperate with the ecosystem. Why would it be worth their time and effort? What could they gain from such involvement? The following are some of the significant benefits.

1. Enhanced Threat Intelligence and Information Sharing:

a. Consortia often serve as hubs for sharing real-time information about emerging threats, vulnerabilities, and attack techniques specific to the energy sector.
   b. Members can gain access to threat intelligence feeds, security alerts, and analysis from other members and security experts.
   c. This collective intelligence enables the proactive identification and mitigation of potential risks before they can be exploited. For example, a consortium might share indicators of compromise (IOCs) related to a recent phishing campaign targeting energy companies.

2. Access to Specialized Knowledge and Expertise:
   a. Consortia bring together cybersecurity professionals with diverse skills and experience relevant to the energy industry.
   b. Members can learn about BP, innovative security solutions, and effective strategies for protecting OT and IT environments.
   c. This access to specialized knowledge can be invaluable for companies that may lack in-house expertise in specific areas of energy cybersecurity. For example, a company new to smart grid technology can benefit from the experience of others in the consortium who have already addressed its unique security challenges.

3. Improved Incident Response Capabilities:
   a. Consortia can facilitate collaboration during and after cyber incidents.
   b. Members can share lessons learned from past incidents, access shared incident response resources, and coordinate responses to widespread threats.
   c. Practicing incident response scenarios within a consortium environment can help companies improve their preparedness and minimize the impact of an actual attack. For example, a consortium might conduct joint exercises simulating a ransomware attack on a power generation facility.

4. Cost Reduction through Shared Resources and Services:

a. Some consortia may offer shared cybersecurity services, tools, or platforms at a reduced cost compared to individual procurement.
b. Collaborative research and development efforts within a consortium can also lower the financial burden of innovation in cybersecurity.
c. For example, Joint procurement of threat intelligence services or participation in shared security operations centers (SOCs) are examples of potential cost-saving benefits.

5. Stronger Advocacy and Influence:
   a. A united consortium has a stronger voice when engaging with government agencies, regulatory bodies, and technology vendors on cybersecurity issues.
   b. Members can collectively advocate for policies, standards, and solutions that better address the unique challenges of the energy sector.
   c. This collective influence can help shape a more secure and resilient energy ecosystem. For example, a consortium might lobby for specific cybersecurity requirements in new energy infrastructure regulations.
6. Enhanced Reputation and Trust:
   a. Membership in a reputable energy cybersecurity consortium can enhance a company's reputation and demonstrate its commitment to security.
   b. This can build trust with customers, partners, and stakeholders, which is increasingly important in the face of growing cyber threats.
   c. Being part of a recognized consortium signals that the company takes cybersecurity seriously and is proactive in addressing industry-specific risks.

## 2.2 Applicability

This document applies to all organizations in the energy sector ecosystem (Figure 03).

## 2.3 Obligations of the Energy Sector Ecosystem's Organizations

All organizations and companies that are part of the ecosystem (mentioned in Section 2.1 and shown in Figure 03) must fulfill the following requirements:

1. Comply with the proposed framework.
2. Prioritize cybersecurity and take all necessary steps to protect and manage their systems and data.
3. Promote compliance and maintain a good standing, i.e., obtain and maintain certification and compliance in accordance with international standards, implement frameworks, uphold BPs, and comply with state laws pertaining to cybersecurity.
4. Continuously improve organizational cyber and information security capabilities, policies, and procedures. The same should apply to quality assurance aspects.
5. Enhance the organization’s resilience to operational disruptions resulting from cyberattacks.
6. Continuously assess IT, OT, cyber, and information security risks from a systemic, organization-wide perspective within the frameworks of the regulations, directives, and standards to which the organization must adhere.
7. Continuously assess, document, and address cyber and information security risks relevant to the organization's operations and implement mitigation measures.
8. Explore and address cyber and information security scenarios that could affect the organization’s activities and those of its customers, suppliers, and service providers.
9. Explore the scope of potential cyber and information security threats, along with the necessary security measures to address them.
10. Ensure that the methodology for managing and handling cyber and information security events complies with relevant international standards.

# Part I – Non-Technological SbD Issues

## 3. Governance

The boards of directors and senior management of energy sector organizations will play a significant role in achieving the goal of SbD across the sector and ensuring its sustainability throughout the ecosystem.

SbD should be implemented in two situations. The first is when creating a new product or service or changing an existing ICS system. In such cases, SbD should be implemented from the beginning. The second situation is when the organization integrates SbD into an existing ICS environment.

In both cases, the SbD process should be led by the organization's board of directors and senior management. This process begins with the board and senior management adopting a cybersecurity mindset and implementing cybersecurity methods and policies. The subsections that follow describe the relevant responsibilities and activities of the board, senior management, chief information security officer (CISO), and the internal IT audit function.

### 3.1 Board of Directors

We make the following assumptions about the board's responsibilities regarding SbD:

- As part of its managerial activity, the board of directors affirms the organization's commitment to cyber and information security and approves its cyber and information security risk management strategy. The board lays the groundwork for the SbD framework's implementation across the organization. It influences the organization's behavior regarding the adoption of the framework as a whole.
- The board's cyber and information security actions, risk management, outsourcing, backup and recovery from cyber incidents and disasters, etc., are aligned with international standards, frameworks, BPs, and state regulations.

Given those assumptions, the board is responsible for:

1. Approving the organization's SbD strategy and policies.
2. Appointing a subcommittee of the board of directors to oversee the implementation of the SbD framework in all organizational OT, IT, environments, computers, networks, systems, and maintenance projects.
3. Ensuring that all annual and interim work plans regarding cyber and information security, business continuity, and disaster recovery are aligned with the SbD framework.
4. Supporting inter-organizational collaboration in the implementation of the SbD framework.
5. Ensuring that all OT controls, IT controls, and controls at the interface between OT and IT are aligned with the SbD framework.
6. Ensuring that the budget allocated for SbD activities is sufficient.
7. Ensuring that there are no obstacles preventing the implementation of the organization's SbD strategy.

Note that in the absence of a board of directors, these responsibilities will be carried out by the organization's senior management.

## 3.2 Senior Management

We make the following assumption about the senior management's responsibilities regarding SbD:

1. Like the board of directors, as part of its managerial activities, senior management affirms the organization's commitment to cyber and information security, risk management, outsourcing, backup and recovery from cyber incidents and disasters, etc.
2. It ensures that all organizations' activities, as mentioned in clause 1, are aligned with international standards, frameworks, BPs, and state regulations.
3. Allocate the necessary resources to perform all that is required to implement the activities, as mentioned in clause 1, including all security measures necessary to support auditing processes and ensure that adequate training is provided to the relevant unit.

Given those assumptions, the organization's senior management is responsible for:

1. Establishing an organizational mindset for SbD management and supervising its implementation and ongoing maintenance.
2. Establishing the organization's SbD strategy and approve its policies.
3. Ensuring that cyber and information security policies are coordinated and aligned with the SbD framework.
4. Discussing, at least on an annual basis, whether the organizational SbD policies need to be modified.
5. Allocating the necessary resources for the implementation and ongoing maintenance of the SbD framework and policies.
6. Reviewing and discussing reports (regularly scheduled, ad-hoc, or required) regarding implementation of the SbD framework.
7. Reviewing and discussing monthly reports on significant cyber and information security incidents and analyzing their implications for the organization. As mentioned in Section 1, these reports and discussions may influence the SbD framework.
8. Establishing an SbD steering committee headed by a member of the organization's senior management.
9. Designating a CISO for the organization and defining his/her scope of responsibilities and areas of authority, with the expectation that he/she adopts a proactive approach to cyber and information security and SbD.
10. Ensure its cyber and information security safeguards are certified and comply with relevant laws, regulations, standards, and BPs.
11. Promoting inter-organizational collaboration to support the implementation of the SbD framework.

## 3.3 Internal Audits

The internal audit function is an organizational unit responsible for all audit activities, including operational, security, compliance, environmental, IT, and OT audits, within an organization.

We make the following assumptions about the internal audit unit's responsibilities regarding SbD:

- The internal audit activities include auditing cyber and information security, OT and IT environments, communication, risk management, outsourcing, survivability, backup, and recovery from cyber incidents and disasters, among other areas.
- The internal audit unit activities are aligned with international audit standards, frameworks, BPs, and state regulations.

Given those assumptions, the internal audit is responsible for:

1. Auditing the processes associated with the implementation and ongoing maintenance of the SbD framework.
2. Ensuring that the unit receives training to improve its capabilities and performance.
3. Auditing of all aspects of the SbD framework’s implementation and ongoing maintenance at least once a year. Furthermore, a thorough audit should assess the organization’s survivability, backup, and recovery processes related to the SbD framework.
4. Reporting the SbD audit findings to the organization's senior management and board of directors, who will discuss them as stated in the previous subsection.

## 3.4 The Chief Information Security Officer

The appointment of a CISO, the definition of this position's responsibilities, particularly in relation to SbD, and its place in the organizational hierarchy are assumed to have been performed by senior management (as mentioned in Section 3.2) in accordance with international standards, frameworks, BPs, and state regulations.

Given those assumptions, the CISO is responsible for:

1. Advising the board and senior management on cyber and information security management in the OT and IT environments and the actions needed to fulfill the requirements of the SbD framework.
2. Establishing and formulating the methodology, policies, and procedures required to implement the SbD framework in the organization. The SbD methodology and policies should be approved by the board and senior management (as stated in Sections 3.1 and 3.2).
3. Coordinating and integrating all organizational efforts to implement the SbD framework.
4. Ensuring that the SbD framework, policies, and procedures are properly implemented.
5. Ensuring that the relevant organizational units are engaged in a continuous process of evaluating cyber and information security risks, understanding the impact of their activities on the implementation of the SbD framework.
6. Establish reporting procedures to ensure that the relevant organizational units submit ongoing and ad-hoc reports on the implementation of the SbD framework and review those reports.
7. Evaluating the effectiveness of the SbD framework's implementation in cyber and information security readiness training, such as internal exercises and drills with business partners.
8. Enhancing organizational awareness regarding the concept of SbD and the activities required for its implementation.
9. Analyzing cybersecurity and information security incidents worldwide, assessing their potential impact on the organization, including the implementation of the SbD framework, and taking any necessary measures to mitigate that impact.
10. Reviewing and assessing the security controls in the organization on a regular and ad-hoc basis.

11. Being aware of all regulations, standards, and BPs related to security and quality assurance that the organization should adhere to and validating suppliers, business partners, integrators, HW and SW manufacturers, etc., follow the relevant regulations, standards, and BPs, including accreditation and certification.
12. Ensure that cyber and information security safeguards are certified and comply with relevant laws, regulations, standards, and BPs.
13. Preparing and submitting semi-annual reports on activities related to the SbD framework to the board and senior management. These reports will describe the organization's cyber and information security defense level, including weaknesses and vulnerabilities, as well as available countermeasures, along with the activities and budgets required to enhance organizational defenses.
14. Collaborating with relevant institutions involved in SbD issues.

### 3.4.1 Appointment

1. The CISO will be appointed by senior management and tasked with specific responsibilities, duties, and organizational authority.
2. The CISO should possess the appropriate background (i.e., education), skills, and experience in cyber and information security for both OT and IT environments.
3. The CISO will serve as the decision-maker in cases of conflicts, misunderstandings, disagreements, and other issues regarding the organization's obligations to fulfill the SbD framework’s requirements.
4. The CISO should not hold any other position or have other responsibilities within the organization, as this could impact his/her role as CISO.
5. The organization's senior management will provide the CISO with all the necessary resources to fulfill their responsibilities.

### 3.4.2 CISO’s Place within the Organizational Hierarchy

1. A member of the senior management will directly supervise the CISO.

2. The CISO will have the authority to choose the necessary work and reporting interfaces with other organizational units within the organization, as well as with external organizations. Senior management must approve these interfaces.

# 4. Part II – Technological SbD Issues

In the subsections that follow, the technical requirements for implementing the SbD framework are discussed.

## 4.1 Standards, Best Practices, and Accreditation

This section presents standards, frameworks, and BPs related to ICS security, IT security, and quality assurance. In addition to adopting the standards, frameworks, and BPs listed below, each organization in the energy sector's ecosystem should fulfill the following requirements:

1. The organization should be appropriately accredited. Accreditation is crucial, since it establishes a basis for the organization's level of security.
2. The accreditation should be renewed annually or in accordance with the requirements of the accrediting institution issuing the standard or BP.
3. If a standard or BP does not have a formal accreditation process, an external expert in the subject matter should conduct an accreditation-like process and present the results to the CISO, board, and senior management.

The expectation that all organizations in the ecosystem will be appropriately accredited does not exempt an organization from performing its own on-premises tests for COTS products and services provided by accredited third-party services.

### 4.1.1 List of Standards, Best Practices, and Frameworks

Organizations in the ecosystem should adhere to the standards and BPs as presented in Table 1. (note that this list only contains significant standards; we do not include industry-specific standards, such as IEC 61850—standard for the design of electrical substation automation, IEEE 2030—smart grid interoperability series of standards, and CIS critical security controls, which can be considered best practices).

Table 1: List of standards and BPs (note that the list does not contain all of the standards an organization might follow and adhere to).

| | |
|---|---|
| **General security standards** | |
| ISO27k | set of security management standards |
| NIST CSF | NIST cybersecurity framework |
| NIST SP 800-53 | Security and Privacy Controls for Information Systems and Organizations |
| NIST SP 800-171 | Protecting Controlled Unclassified Information (CUI) in Nonfederal Systems and Organizations |
| **Industrial control system standards** | |
| NIST SP 800-82 | Guide to Operational Technology (OT) Security |
| ISA/IEC 62443 | industrial automation and control system security series of standards |
| IEC 61850 | communication networks and systems for power utility automation |
| FERC CIP | set of standards for protecting critical infrastructure in the energy sector |
| **SDLC standards** | |
| ISO/IEC/IEEE 15288 | Systems and Software Engineering - System Life Cycle Processes |
| NIST 800-218 | Secure Software Development Framework (SSDF) version 1.1 |
| **Quality assurance and quality control standards** | |
| ISO/IEC 25K | set of standards for system and software quality requirements and evaluation (SQuaRE) |
| ISO 9000 | set of quality management standards |
| ISO/IEC/IEEE 29119 | Set of software testing standards |
| **Management and governance** | |
| COBIT | Control Objectives for Information and Related Technology |
| C2M2 | cybersecurity capability maturity model specifically for the energy sector |
| **Related standards** | |
| ISO 55000 | Asset management – Vocabulary overview and principles |
| ISO/IEC 20246 | Software and systems engineering — Work product reviews |
| IEEE 1547 | For Interconnection and Interoperability of Distributed Energy Resources with Electric Systems |
| IEC 62351 | set of standards called "security in power systems and control systems" |
| ISO/IEC 15408 | Information security, cybersecurity, and privacy protection — Evaluation criteria for IT security |

Organizations and suppliers should be certified to meet the standards outlined in Table 1. Certification should always be in accordance with the latest version of a standard or a new standard if the previous one has been replaced. Please refer to Section 4.2, clause 3, for standards that do not require certification or accreditation processes.

The MITRE ATT&CK knowledge base (KB) should be used to meet specific security requirements.

## 4.2 Asset Management

This section outlines the organization's responsibilities for its OT and IT assets. An asset is defined as any data or information, HW, SW, communications equipment, process, policy, or procedure used by the organization. Proper asset management contributes to various aspects of organizational cybersecurity, including visibility, risk analysis, risk management, patch management, event handling, and quality assurance, all of which are building blocks of SbD.

The following subsections outline the asset requirements for each organization, with a focus on inventory, ownership, and other relevant factors.

### 4.2.1 Inventory

1. The organization must document its assets and maintain an up-to-date inventory of them.

2. The asset inventory should be computerized.
3. At a minimum, the asset inventory should include the following assets:
    a. Tangible assets (e.g., HW components) and intangible assets (e.g., SW) are located on the organization's premises.
    b. Tangible and intangible assets that are not located on the organization's premises but are under its responsibility.
    c. Tangible and intangible assets that are not directly under the organization's control but are required for its operation. Such assets could negatively affect the organization when in short supply or malfunctioning. For example, a shortage of third-party cybersecurity experts could delay security testing, preventing the timely implementation of an OT and IT product or service.

### 4.2.2 Ownership

1. Each asset should have a designated owner.
2. The owners of assets are the managers of organizational units within the organization.
3. The owners are responsible for activities such as:
    a. The asset’s daily use.
    b. Ensuring that the asset is properly maintained.
    c. Ensuring compliance with the organization’s policies and procedures regarding the asset's maintenance, development, backup, recovery, and disposal.
4. All cybersecurity-related activities should fall under the direct guidance and supervision of the CISO and their staff.

### 4.2.3 Acceptable Use

1. The acceptable use of an asset is defined as its utilization solely for work assignments.
2. Assets assigned to users or organizational units can only be used for their designated work assignments, subject to their respective roles and authorizations.
3. A user should not disrupt or damage an asset.

### 4.2.4 Asset Mapping and Classification

1. Every asset must have a unique identification.
2. At a minimum, the asset inventory should include the following information for each asset:
   a. The metadata describes the asset (e.g., unique identification number, physical location, owner).
   b. Mapping that enables the organization to fully understand and document its assets in terms of how they relate to each other, not just what they are, but how they work together, affect one another, and how they're connected. The following mappings are required:
      i. A mapping of the logical relations. I.e., understanding how assets interact or depend on each other in terms of functions and data flow (e.g., a database server and an application server may not be directly wired together, but the application depends on the database to function).
      ii. A mapping of physical connections. I.e., understanding how assets are physically wired or connected, including their physical location (e.g., a sensor is physically wired to a gateway device, a PLC connected to a switch).
      iii. Dependencies between assets. I.e., functional dependencies (e.g., a security camera depends on a storage server to save footage; if the server goes down, the camera may not be able to record).
      iv. Constraints. i.e., operational limitations, usage restrictions, and availability constraints (e.g., a network switch is being upgraded, which may temporarily render all devices connected to it unusable; the asset cannot be used during holidays).
      v. Processes using the asset. I.e., an operational or business process that uses the asset (e.g., a valve is used in an oil pump, and that process depends on the actuator to function correctly).

vi. Redundancy and failure. i.e., assets under maintenance, or disaster recovery situations (e.g., two pumps where one can operate while the other is undergoing maintenance).

   c. Asset sensitivity (e.g., patent information).
   d. Cybersecurity measures are implemented to protect the asset.

3. At a minimum, each item in the asset inventory should be assessed in terms of its:
   a. Business importance, i.e., the asset's value to the organization.
   b. The significance in the OT or IT environment, such as the processes that would be impacted in the event of a failure.
   c. Susceptibility to cyber threats.
   d. Potential legal issues might arise if the asset is compromised, tampered with, destroyed, etc.
   e. The potential damage caused to the organization if the asset is compromised, tampered with, destroyed, etc.
4. The asset inventory and its contents should be continuously updated.
5. The categories used to classify assets in the asset inventory should be reviewed annually and whenever changes occur in the OT and IT environments.

## 4.3 Security Infrastructure Requirements

This section presents the fundamental cybersecurity infrastructure requirements associated with SbD.

1. Cybersecurity systems, measures, and safeguards should be determined based on risk assessment and analysis, the sensitivity of the asset, loss prediction, applicable laws and regulations, and applicable standards and BPs.
2. Cybersecurity systems, measures, and safeguards should be continuously updated to the highest level possible in order to properly address the issues mentioned in clause 1 above.

3. Cybersecurity systems, measures, and safeguards should be designed, built, and implemented with “fault-tolerant,” “fail-secure,” “self-healing,” and “self-recovery” capabilities.
4. The implementation, development, and maintenance of cybersecurity systems, measures, and safeguards should not endanger or harm the capabilities described in clause 3 above.
5. Full visibility of all systems, activities, and data should be enabled to detect malicious activities and assist forensic investigations. Additionally, this support should extend to in-house OT, IT, and outsourced activities (e.g., cloud services).
6. An attempt to modify, distort, tamper with, or perform any other activity that was not considered by the system’s designers when designing the cybersecurity system, measure, or safeguard should trigger an alert in the central security incident and event management/security operations center (SIEM/SOC) as soon as it is detected. Regardless of whether the activity causing the alert is performed by an authorized or unauthorized entity, e.g., an employee, SW application, or attacker, the alert should be raised.
7. All cyber and information security safeguards should be certified and comply with relevant laws, regulations, standards, and BPs.
8. Outsourced activities and services (e.g., Internet services, code writing, and cloud services) should comply with the organizational security policies and procedures to ensure that the highest possible level of security is maintained. This means ensuring that the organization is protected from unauthorized access, theft, and loss.
9. Cybersecurity issues should be coordinated with the organizational units responsible for OT and IT development and operations.
10. The organization should implement and manage security safeguards for its OT, IT, data resources, and communication, including but not limited to those described in the subsections below.

### 4.3.1 Access Controls

1. A mechanism for identification and authentication should be in place to enable access to the organization's OT and IT environments, including production, test, and development environments.
2. Multi-factor authentication (MFA), such as fingerprints and a strong password, should be required to access the environments mentioned in clause 1.
3. A mechanism for identification and authentication should be in place to access the HW components.
4. Dynamic access control using a ZT mechanism should be implemented in the OT and IT environments and between them.
   a. Access to organizational assets should be granted with respect to internal and external attributes.

      **Internal attributes** such as the IP address, physical location of the device requesting access (e.g., on-premises, in a cloud environment, at an employee's home), type of device (e.g., laptop, smartphone, PLC, server), communication method, firmware version, and information regarding Trusted Platform Module (TPM).

      **External attributes** include indicators that the organization is currently under attack and published vulnerabilities relevant to the organization.

### 4.3.2 Remote Access Control

Remote access control provides employees and external companies with secure access to the organization's OT and IT environments for maintenance, inquiries, and repair activities. There are two types of remote access.

The first type involves remote access required for ongoing operations, such as monitoring a remote gas pipeline.

The second type involves remote access required to handle failures or perform repairs. For example, a malfunctioning SW or HW component may require the manufacturer to investigate it remotely.

1. Generally, remote access to organizational resources should be avoided as much as possible. However, if it is unavoidable, it must be conducted in a secure, controlled, and restricted manner.
2. A clear set of criteria and procedures should be established to explain when, how, by whom, and for how long remote access may be used.
3. Remote access to organizational resources, e.g., OT and IT environments, should be implemented using technology that meets, at a minimum, the following requirements:
    a. A specific and dedicated IT component, i.e., a gateway, that possesses at least the following security-related capabilities should be used to provide remote access to the organizational assets:
        - A specific authentication mechanism enables (1) initial authentication of the remote user and (2) continuous authentication of the remote user throughout the session.
        - All communications to and from the remote site should be routed through a specific mechanism. This mechanism will facilitate ending the session after a predefined period or timeout.
        - Encryption of communication between the remote site and the local site.
        - A detailed audit log collecting all communication data should be maintained.
    b. The organization should have two different gateways: one for internal operational communication and another that supports external communication.
    c. The organization should implement technological measures, such as continuous identification and authentication of communication endpoints, to ensure and validate these endpoints.
    d. The organization should use dedicated DMZs for each gateway.
4. Requirements for internal operational remote access, i.e., the first type described above (remote access required for ongoing operations):

a. The gateway for internal operations should be operational.
b. Traffic should be logged in a dedicated audit log file.
c. Traffic should be analyzed for potentially malicious code or activity in real time.
d. Callers should be identified and authenticated using a secondary communication channel to initiate the remote session.
e. Strong MFA (at least 2FA) should be used, e.g., a user code and token that generates random passwords. A certificate mechanism should be used when an HW component requires access.

5. Requirements for external remote access, i.e., the second type described above (remote access required for failure or repair operations performed by an external entity):
    a. By default, the gateway for external remote access is turned off and not operational. It should only be powered on and connected to the network when remote access is required.
    b. Traffic should be logged in a dedicated audit log file.
    c. Traffic should be analyzed for potentially malicious code or activity in real time.
    d. Only the organization should initiate a remote access session.
    e. Callers should be identified and authenticated, using a secondary communication channel to initiate the remote session. Sessions between the external entity and the organization should be initiated manually.
    f. The remote session should be terminated manually or after its predetermined time limit has expired, whichever occurs first.
    g. Access should be protected by strong MFA (at least 2FA), such as a user code and one-time password (OTP).
    h. A parallel communication channel, such as Zoom, should be used during the session.

### 4.3.3 Encryption

The SbD encryption requirements are as follows.

1. Cryptography algorithms and key management must comply with the FIPS 140-2 standard, a similar up-to-date international standard, or another standard superseding the encryption standard in use.
2. The encryption should be performed using the most secure encryption schema.
3. Organizations should use the maximum key length permitted by the cryptography algorithms, such as encryption and digital signatures.
4. Regulations or state laws may stipulate a specific encryption scheme, e.g., AES with a key length of only 128 bits; in such a case, the organization must follow the stipulated encryption scheme using the maximum key length.
5. The organization should use digital signature algorithms recognized as international standards.
6. Encryption and authentication keys should be exchanged according to established international standards and protocols.
7. The encryption keys should be generated and stored in a manner that prevents theft, loss, leakage, or any other risks from occurring anywhere they are stored. Encryption keys should be stored in accordance with international standards such as FIPS 140-2 or the most recent equivalent.
8. Various devices, such as firewalls, routers, PCs, servers, PLCs, and IoT devices that serve as PLCs, should support the following, at a minimum:
   a. The use of symmetric encryption, such as AES, DES, or the equivalent.
   b. The use of asymmetric encryption, such as RSA, ECC, or the equivalent.

### 4.3.4 Certificate Mechanisms

This section presents the SbD requirements regarding certificate mechanisms and certificates issued by certificate authorities, such as RSA.

1. A certificate mechanism should support the highest level of certification known to date.
2. HW and SW manufacturers should be able to support certificates.
3. Certification should be required for HW components, SW, applications, and users.

4. Man-to-machine and machine-to-machine connections should validate the certificate before establishing a connection.
5. An uncertified HW component, SW component, application, or user should not be allowed to connect to or operate in the organization’s OT and IT environments.
6. Certificates should be continuously checked and validated during operation time.
7. Certificates should only be issued by recognized certificate authorities, e.g., RSA.

### 4.3.5 Date and Time Synchronization

Modern IT and OT environments use time synchronization. Some of the reasons are: (1) Incident analysis and troubleshooting. Time synchronization is crucial for correctly ordering events, diagnosing problems, and performing forensic analysis; (2) process control and automation. In OT environments, processes rely on tightly coordinated operations. Synchronization of time among sensors and actuators is crucial for ensuring that actions occur at the correct time; (3) in some OT environments, such as grid control, synchronization is essential for reliability and safety; (4) correlate events across the network [Dang 2023, Yiğitler 2020]. Hence, we ought to request the following:

1. All digital equipment, including IoT and IIoT devices, routers, PCs, and PLCs, should be synchronized with the current date and time.
2. There should be at least millisecond precision in time synchronization.

### 4.3.6 Audit Logs

Organizations should follow NIST special publications or the equivalent for proper audit log implementation, e.g., NIST 800-92 (Guide to Computer Security Log Management). This section presents the audit log requirements relevant to SbD.

1. Log files should be robust and protected against malicious activities (attacks), data manipulation, and deletion by unauthorized entities (such as individuals or parties).
2. The log recording mechanism must be safeguarded against manipulation, subversion, and bypassing to ensure its integrity.

3. Administrators should not be able to modify, subvert, delete, or bypass log records.
4. Audit logs should document (1) all system events, such as logon, logoff, identification, and authentication; (2) access to assets, activities performed on assets; and (3) system responses, including success and failure return codes.
5. The records of local logs stored on a device (i.e., local logs) should be transmitted to a central log. Data transmission (i.e., log records) should be protected against malicious attacks, data manipulation, and data deletion by unauthorized parties.
6. Log information should, at a minimum, consist of the following:
   a. Date and time (to the millisecond).
   b. Identification of the transaction, event (e.g., transaction ID, event ID), etc.
   c. Identification of the HW and SW components (e.g., certificates), user ID, communication addresses (e.g., source and destination addresses), etc., is essential. Please note that the identifiers of the above components should be unique within the organization.
   d. Any other information relevant to the transaction or event that may facilitate forensic investigation.
7. The log fields should be parameterized, and the CISO should determine which fields appear in the log record.
8. The activities of the organization's senior management, including the CISO, should be logged.
9. It should not be possible to manipulate, subvert, delete, or bypass the log recording mechanism.

#### 4.3.7 Data Integrity

This section emphasizes the critical importance of implementing mechanisms to ensure data integrity across all systems. In this context, the term "data" encompasses all forms of information and code within Information Technology (IT) and Operational Technology (OT) environments. This includes, but is not limited to, database tables, configuration parameters, communication messages, source code, and executable code. These considerations apply to all operational stages—development, testing, and production.

To uphold data integrity, the following principles should be observed:

1. Data integrity must be preserved across all states of data: while in transit (data in motion), when stored (data at rest), and during processing (data in use).
2. Dedicated mechanisms, such as digital signatures or cryptographic hashes, should be employed to enforce data integrity and detect tampering.
3. Integrity validation should be performed continuously, leveraging automated tools and monitoring systems to identify unauthorized modifications.
4. Any detected deviations from expected integrity should be immediately reported to the SIEM/SOC for investigation and response.

#### 4.3.7.1 Data in Motion Integrity Mechanisms

The following requirements must be implemented for data in motion:

1. Data in motion should be encrypted and digitally signed. This includes data transmitted within the organization's premises and between the organization and the external environment.
2. The recipient of the data should check the digital signature.
3. Data in motion should be protected against malicious attacks, manipulation, and deletion by unauthorized parties.
4. Data should be verified prior to encryption to ensure accuracy and check for malicious code. The same verification/checking process should be performed after the decryption process.
5. Data should only be transferred among predefined nodes. The transfer mechanism should support parameters including, but not limited to, (1) data identification (e.g., information, code), (2) direction of movement, (3) statistics (such as the number of bytes being transferred), (4) date and time of the transfer, and (5) allowed routes of the data movements.

#### 4.3.7.2 Data at Rest Integrity Mechanisms

The following requirements must be implemented for data at rest:

1. Data at rest should, by default, be encrypted.
2. The CISO determines whether to change the requirement in clause 1. Any changes should be made following a rigorous investigation that examines aspects such as the data's sensitivity, information lifespan, and use. The organization's senior management should approve those changes.
3. The organization should implement a specific mechanism for identifying and preventing data integrity damage. Any integrity deviation should be reported immediately to the SIEM/SOC.

#### 4.3.7.3 Data in Use Integrity Mechanisms

The following mechanism should be implemented for the data in use.

1. Data in use should be protected in memory (e.g., RAM, ROM, EPROM) and the CPU cache memory.
2. The protection mentioned in clause 1 should include the prevention of unauthorized modification, distortion, or reading of the data.
3. As described in clause 2, any unauthorized activity should be reported to the SIEM /SOC. If such an event is identified, it is up to the CISO to determine the response.
4. When data in memory is encrypted, it must be decrypted. Therefore, as stated in clause 2, the decryption of data in use should take place in a protected area (e.g., CPU cache memory).

### 4.3.8 Non-Repudiation

Dedicated means should be used to enforce non-repudiation, such as digital signatures using public-key cryptography, timestamps, secure hash algorithms, secure messaging protocols, or more advanced technologies, such as a trusted third-party timestamping authority, blockchain, or zero-knowledge proof protocols.

#### 4.3.9 Physical Security

Below, we list the requirements related to the physical security of the OT and IT environments.

1. Physical security should be provided at all organizational sites.
2. The term "organizational sites" includes, but is not limited to, buildings, offices, operational sites, IT sites, OT sites, communication sites, backup sites (such as hot, warm, or cold backup sites), and sites where the organization's data resides.
3. We assume that comprehensive physical security is in place at all locations. Note that weak physical security may impact the organization's implementation of SbD; this topic is outside the scope of this document.

#### 4.3.10 Maintenance

Although maintenance should be an ongoing process, it does not appear to be frequently performed in the ICS environment, particularly in the OT environment. Therefore, an organization may find itself with an OT device susceptible to cyberattacks.

We address the maintenance process requirements in Section 5.6.

#### 4.3.11 Risk Assessment and Audit Activities

Cyber risk assessment and audit activities identify, analyze, and evaluate the risk associated with an organization's cybersecurity strategy and implementation.

1. Risk assessment should be a continuous process, performed with respect to:
   a. In-house organizational changes in HW, SW, communication systems, etc.
   b. Changes that occur externally, including changes in service providers, customers, business partners, published information about cyberattacks, vulnerabilities, and academic research.
2. The organization should consistently conduct vulnerability audit scans across all elements of its networks. For instance, it should implement a mechanism that scans all communication devices, PCs, servers, and more for (1) vulnerabilities, (2) missing patches, (3) end-of-life hardware and software, and so on. The vulnerability audit scans should encompass all ICS environments, including both OT and IT.

3. The organization should perform quarterly penetration and robustness tests in IT and OT environments. Usually, these networks are large, so covering all of them quarterly may not be possible. In such cases, the entire network should be examined at least annually.

### 4.3.12 SIEM/SOC

An organization's security incident and event management (SIEM) system, along with its security operations center (SOC), identifies data breaches, alerts the organization to potential cyber events, and enables a prompt response to them.

1. Every organization should have a SIEM/SOC in place.
2. The SIEM/SOC should serve both the IT and OT environments, integrating information from each.
3. Incident and event data should be transferred to the SIEM/SOC for analysis and recommendations to determine how the organization should react to the incident/event.

It is highly recommended that an industry-wide SIEM/SOC be established for all members of the energy sector ecosystem, in addition to the organization's SIEM/SOC. This entity will gather and analyze the information from the member organizations' SIEM/SOCs.

## 4.4 Architectural Elements

This document's objective is to establish the requirements for implementing the SbD framework. Basic cybersecurity requirements should also be met to achieve the goal of implementing SbD in ICSs. This section outlines the fundamental architectural security requirements for ICS, including its communication networks. Note that some information presented in this section has already been discussed in previous sections.

1. The organization should implement a secure physical and logical architecture in order to prevent, detect, rectify, and recover from breaches in its OT, IT, and communication systems.
2. The organization should implement methods and mechanisms to prevent data breaches.
3. The organization should continuously seek to enhance its systems' cybersecurity measures and policies.
4. The requirements for network security elements can be found in Section 4.5.
5. The requirements for work environments can be found in Section 5.2.
6. Authentication and authorization mechanisms should be used to enable access to the organization's network and systems. Human authentication should be based on MFA, and device authentication should be based on strong digital certificates.
7. The organization should prohibit the use of portable media by default. In extreme cases, portable media may be used, but only after it has been checked for malicious content.
8. The organization should use anti-malware and sanitization mechanisms to prevent malicious content, such as malware, from infecting its network, systems, and servers.
9. The organization should use anti-malware and sanitization mechanisms to prevent incoming and outgoing files containing malicious content from infecting its network and external networks.
10. The organization should use encryption methods to protect its data files. Permission to inspect the data files should be given on a need-to-know basis and for audit purposes.
11. The organization should monitor its internal networks and systems, such as PCs and servers, for malicious activity and policy violations by internal and external parties.
12. The organization should strive to implement cutting-edge technology based on anomaly detection algorithms in order to detect zero-day attacks, malware, and other malicious activities.
13. Each device should be equipped with a TPM, which will be used in activities such as the boot process and memory control.

14. Each device in levels 0-3 of the Purdue model should be duplicated. Ideally, this will enable proper and timely maintenance activities to be performed on level 0-3 devices. Note that IT device duplication is a common practice in online environments (e.g., Internet sites).
15. A level 0 OT device should be connected to at least two level 1 OT devices. This requirement could help in (1) identifying attacks on level 0 OT devices and (2) responding to the attack.

## 4.5 Network

In this section, we discuss the cybersecurity of network components. It is assumed that the organization has implemented safeguards such as firewalls, intrusion detection and prevention systems (IDPSs), identification and access control systems, data leakage prevention (DLP) systems, and other wired and over-the-air communication protection mechanisms.

### 4.5.1 Areas of Emphasis

In addition to the network requirements mentioned later in this section, we wish to emphasize the following:

1. Unused ports in the network switches and routers should be disabled; this can be done manually or by using an SW application that controls access to network resources.
2. Communication components should be hardened to an adequate level, in accordance with standards and cybersecurity best practices.
3. The configuration of all network elements and cybersecurity measures should be backed up.

#### 4.5.1.1 Network Management

This section deals with basic network management requirements.

1. The organization should formulate policies to enhance the cybersecurity of its network elements, systems, and workstations.
2. A network management system (NMS) should be used to control the OT and IT networks' components, such as routers and bridges.
3. At a minimum, the NMS should have the following properties:
    a. The NMS should be protected by safeguards that monitor and control the incoming and outgoing network traffic based on predetermined security rules (e.g., a firewall, IDPS) and behavioral mechanisms that can detect internal and external attacks, malicious activities, and policy violations.
    b. Depending on the organization's environment, the NMS should be logically divided into different parts.
    c. Administrators should not be permitted to manage the network from their workstations, which may have other connections, such as a direct connection to the Internet.
    d. Administrators should use MFA to access the NMS.
4. All of the organization's administrative activities should be audited, recorded in logs, and sent to the SIEM/SOC.

#### 4.5.1.2 Network Elements

The requirements regarding basic network security settings are listed below.

1. The organization's internal networks should be segmented according to criteria related to the organizational structure, functionality, information sensitivity, and risks.
2. As part of the internal network, i.e., networks inside the organizational premises, security measures should be implemented (e.g., firewalls between the segments, relay mechanisms, information filtering) in order to prevent the injection of hostile code. In addition, proxy systems should be used to conceal the segment infrastructure from hostile actors, and measures should be implemented to identify, detect, and raise alerts regarding suspicious activity, data and data movement, applications, and access.

3. Data movement, access, and security rules related to internal networks should be predefined and monitored for suspicious activities. The monitored data and related activities should be logged and sent to the SIEM/SOC.
4. The organization should maintain strict separation between its internal networks and all entities outside its premises, including the Internet, suppliers, service providers, and customers.
5. The organization should implement security measures to maintain the security of its connections with external parties (e.g., customers, suppliers, and business partners).

### 4.5.2 Network Data Encryption

This section is included for the sake of completeness regarding the network topic.

1. As stipulated in Section 4.3.7.1, data in motion should be encrypted.
2. Management traffic should be encrypted.

## 4.6 Remote Access

This subject is dealt with in Section 4.3.2.

# 5. Security by Design for System Development

## 5.1 Introduction

In this section, we discuss the general aspects of SbD requirements related to the development or acquisition of HW or SW components for OT and IT environments.

In this context, the term “artifacts” (or "products") refers to any HW, SW, or combination of the two.

Three types of artifacts are considered.

a. An artifact developed by the organization on its own.
b. An artifact purchased by the organization.
c. An artifact that is a combination of the two types of artifacts mentioned above (e.g., using an API developed by the organization in a purchased application)

We do not propose a new SW, HW, system, procurement, or development lifecycle; instead, we present cybersecurity concerns related to system procurement and development that influence the SbD concept. We believe that adhering to these requirements will result in a higher level of cybersecurity and robustness against cyberattacks.

1. All artifacts should be certified according to relevant standards, including cyber and information security and quality assurance standards.
2. All information regarding internally developed artifacts must be documented. The documentation should include details about activities during the development of lifecycles, internal and external testing, and certification processes. It is essential to maintain the documentation to ensure it remains complete and accurate (see note below).
3. The manufacturer or creator of an artifact should provide complete and accurate documentation to ensure its compliance with the required standards, BPs, regulations, and testing, including security testing (see note below).

4. SW development should be performed using modeling languages, such as the Unified Modeling Language (UML) or Systems Modeling Language (SysML), since modeling languages can support the entire SW development lifecycle (SDLC).
5. There should be complete bidirectional traceability in the SW development process.
6. Traceability information should be documented and logged.
    a. In the development and maintenance phases, traceability should be ensured by linking artifacts to their requirements, design, coding, testing, and documentation. This information should be highly protected.
    b. Traceability at runtime should track every change made to a data item, log its movement between modules, APIs, systems, etc., and log all information about the change.
7. The system should support full visibility to all of its components.

Please note that many standards require documentation of the activities outlined in the standard document. Therefore, we recommend including knowledgeable technical writers as part of the development staff. The same applies when purchasing an artifact or integrating a system.

## 5.2 Work Environments

It is essential to create distinct work environments for development, testing, and production. Some reasons for this include:

- Establishing different work environments prevents the development and testing teams from accidentally corrupting the code in the production environment.
- It also enables parallel development, testing, and maintenance activities.
- By protecting the integrity of the production data and code, establishing different work environments improves security.
- The development and testing environments can be viewed as a line of defense, enabling the identification of bugs and security vulnerabilities before the code is moved to the production environment.

The rest of this section presents the requirements imposed on the organization regarding the work environment.

1. At a minimum, an organization should define the following work environments: development, testing, near-production, and production, i.e., operations. It is up to the organization to determine whether any other environment is necessary.
2. Generally, all organizational environments should be isolated from the outside world, including third parties.
3. Separation of the environments should be strictly enforced.
4. There should be a high level of cybersecurity protection for all environments.
5. Data movement (such as new SW, code updates, or firmware updates) between the environments should be performed through a specific mechanism and coordinated by all participants.
6. Concerning the mechanism mentioned in clause 5:
   a. Data that moves between two environments should be analyzed for malicious code.
   b. An audit trail should be maintained for each instance of data movement.
7. Development teams should not be permitted to access the production environment. However, if such access is required, for example, in the event of a failure, the development teams should handle it within the production environment. Remote access should only be permitted when specific, stringent measures are in place, as stipulated in Section 4.3.2. Each access permission and activity that allows a development team to access the production environment should be accompanied by an audit trail.
8. Security tests, including penetration and robustness tests, should be performed on all possible attack scenarios and access paths, including the remote access path.

## 5.3 Lifecycle

This section provides a general description of lifecycle activities that are to SbD. We do not delve deeply into each aspect of the lifecycle, as there are numerous books, papers, and BPs available for this purpose. Instead, we focus on lifecycle-related cybersecurity and SbD requirements. In this context, the general lifecycle consists of the following phases: the requirement, design, coding or development, testing, deployment to production, maintenance and replacement, and disposal phases.

### 5.3.1 Requirement Phase

In this phase, relevant aspects of security requirements should be identified as part of the risk analysis process to the greatest extent possible. Regarding security, we do not differentiate between an artifact (e.g., HW, SW) developed or purchased by an organization. Consequently, organizations should create a detailed questionnaire concerning relevant aspects of security and either develop artifacts or consider purchasing artifacts based on the outcomes of the questionnaire.

Generally, the requirement phase includes at least the following items to be considered:

a. Risk assessment and analysis
b. Quality assurance
c. Identification and authentication
d. Access control
e. Logs/historians
f. Audits
g. Tests
h. Security
i. Communication
j. Remote access
k. Segregation of duties
l. Segmentation
m. Perimeter security
n. Backup and recovery
o. Physical security
p. Law and regulations

This document does not discuss each of the aforementioned items in detail. Instead, it examines several activities related to achieving SbD.

The requirement phase typically occurs in cycles [Pohl 2016]. It begins with a high-level risk assessment and analysis. As we move through the requirement phase, we conduct a risk analysis to identify any uncovered risks and security vulnerabilities [Ardi 2023].

#### 5.3.1.1 Risk Analysis – Understanding the Risks

In this document, we do not aim to explain how to perform risk assessment or analysis. Instead, we want to emphasize the importance of conducting risk assessments and analysis in any internal development project or when purchasing an external HW and SW product.

Organizations and manufacturers should perform risk assessment and analysis before starting a project (e.g., internal development or purchasing). This process will contribute to improved understanding regarding the security risks and aspects of security that should be considered before embarking on the project. Note that in the context of this document, risk assessment and analysis do not refer to non-security-related project risks (e.g., adherence to the timetable and budget).

The outcome of the risk assessment and analysis process may include:

- An understanding of the risks the organization may face in relation to the artifact developed or purchased.
- An agreement between the participants (e.g., users, developers, management) regarding the security risks and mitigation activities pertaining to the artifact developed or purchased.
- A high-level map of all security requirements and threat models pertaining to the artifact developed or purchased.
- A document stating all of the security requirements.
- A repository or KB that includes all information about the security requirements, which should be updated at every stage of the development process.

- An understanding of the interdependencies between various security requirements.
- A thorough understanding of the extent to which the organization is capable of meeting the security requirements.
- An overview of how to validate and verify the security requirements.
- Prioritization of the security requirements.

Whenever there is a change in an artifact (HW or SW), infrastructure, or security requirement (e.g., a new regulation), a new risk assessment and analysis of the security requirements should be conducted to ensure full coverage of the security risks.

#### 5.3.1.2 Quality Assurance Requirements

QA and security are about reducing risks. Several studies have found that security and QA are correlated, i.e., applications with many QA issues tend to have more security issues [Arzt 2021, Woody 2014].

It is challenging to develop high-quality products (i.e., products that have minimal faults and are reliable, accurate, easy to maintain, secure, and robust to cyberattacks), whether they are HW, SW, or systems. Users perceive high-quality products as trustworthy, and thus, they will prefer such products or systems to those that frequently fail, are difficult to maintain and unreliable, or are not secure (e.g., prone to information leakage, easily exploitable), which are perceived as untrustworthy. Fundamentally, security is all about trust [Granger S. 2001]; therefore, to attain a high level of security, attention must also be given to QA, which affects security aspects such as reliability and availability, hence influences the implementation of SbD [Mallouli 2022, ISO/IEC 25010:2011, Yeboah-Ofori 2020]. A comprehensive QA program that addresses all aspects of QA is examined and has the potential to reduce security vulnerabilities and deficiencies.

The QA requirements are listed below.

1. Any manufacturer or organization that supplies SW or HW artifacts to an organization in the energy sector ecosystem should adopt QA and quality management standards such as the ISO/IEC 250* stack of QA standrds, e.g., ISO/IEC 25022:2016, ISO/IEC

25023:2016, ISO/IEC 25030:2019, ISO/IEC 25065:2019, or any other standard that supersedes them.

Table 2: The table below presents the quality requirements stated in ISO/IEC 25010. The bold represents a grouping of the requirements.

| | |
|---|---|
| **Functional Suitability** | **Security** |
| Functional completeness | Confidentiality |
| Functional correctness | Integrity |
| Functional appropriateness | Non-repudiation |
| **Compatibility** | Accountability |
| Coexistence | Authenticity |
| Interoperability | **Maintainability** |
| **Compatibility** | Modularity |
| Coexistence | Reusability |
| Interoperability | Analyzability |
| **Usability** | Modifiability |
| Appropriateness recognizability | Testability |
| Learnability | **Portability** |
| Operability | Adaptability |
| User error protection | Insatiability |
| User interface aesthetics | Replaceability |
| Accessibility | **Performance Efficiency** |
| **Reliability** | Time behavior |
| Recoverability | Resource utilization |
| Maturity | Capacity |
| Availability | |
| Fault tolerance | |

Note the following:

1. Several of the quality requirements in the ISO/IEC 250* stack of standards are related to security, such as availability, accountability, and authenticity.
2. The requirements are non-functional. Consequently, they must be translated into functional requirements to be realized.
3. In the literature, the terms dependability, trustworthiness, and survivability are used interchangeably to describe SW security properties [Kahtan 2014]. Therefore, it can be assumed that when an organization declares that its systems are secure, it also declares that its systems are trustworthy and capable of withstanding cyberattacks.

4. ISO/IEC 25010 defines safety as "freedom from risk."

#### 5.3.1.3 Identification and Authentication

Identification and authentication technologies are discussed in standards and BPs. Therefore, we do not discuss them in this document. Please refer to Sections 4.3.1 and 4.3.2 for the identification and authentication requirements.

#### 5.3.1.4 Access Control

Access control technologies are discussed in standards and BPs. Therefore, we do not discuss them in this document. Please refer to Sections 4.4.1 and 4.3.2 for the access control requirements.

#### 5.3.1.5 Logs/Historian

An organization should comply with the requirements outlined in NIST Special Publication 800-92, "Guide to Computer Security Log Management," or any other standard that replaces it.

In addition to the requirements listed in Section 4.3.6, the following points should be emphasized.

1. Transactions and activities in the IT and OT environments should be logged. This information should include, but is not limited to, the following:
    a. Time and date (to the millisecond).
    b. A unique ID of the transaction and event.
    c. A unique ID of the device and user.
    d. Device source and target addresses. This implies that the address and IDs are unique within the organization.

e. Any other information relevant to the transaction or event that may facilitate forensic investigation.

2. Logs should be protected from malicious activities like deletion and record modification.
3. The log records should be sent to a central point for analysis, i.e., the SIEM/SOC.
4. The data fields in the log should be parameterized.
5. Log parameters should only be changed by an authorized user, i.e., an administrator, whose activities should also be logged.

#### 5.3.1.6 Audit

Auditing provides another layer of security control for the protection of an organization's assets and network. Thus, we consider auditing a must for realizing the goal of SbD.

In addition to the requirements listed in Section 4.3.11, the following points should be emphasized.

1. Audit activities can be divided into two major categories.
    a. The first category consists of a methodical inspection, examination, and evaluation of an organization's ICS (OT, IT, and networks), infrastructure, policies, operations, etc. This type of audit is performed by professional auditors. Such audits usually include:
        - An audit of the entire ICS environment (OT, IT). Such an audit should be conducted on an annual basis.
        - Audit activities, such as penetration tests and other non-destructive security audits, should be conducted quarterly or after any significant change in the ICS environment (OT, IT).
    b. The second category consists of near-real-time audits on the activities of all components and system logs/historians. These audits are performed by aggregating all logs (e.g., OT, IT, HW, and communication logs, including SW and system logs), scanning the aggregated log, and analyzing it to identify abnormal activity. Note that this type of auditing is highly dependent on the organization's

visibility requirements, since some of the logs may be encrypted or contain partial information.

2. All audit reports from clause 1 should be distributed to the organization's senior management, directorate, CISO, and other individuals responsible for the audited systems.

#### 5.3.1.7 Testing and Validation

Security testing and validation of HW and SW takes different forms during different phases of the artifact lifecycle (e.g., verifying secure coding during the coding phase, conducting security penetration tests before and after installation in a production environment). As such, there may be a need for additional security controls to protect the organization's assets. In the requirements phase of the lifecycle, the following points must be emphasized.

1. Testing must be comprehensive and include security and non-security tests.
2. A thorough understanding of issues such as: (1) how security tests should be performed, (2) assessing the necessary testing time, required tester expertise, optimal test environment, etc.
3. The non-security tests (e.g., sanity tests) should be performed from a security perspective, considering the ability to subvert the system (e.g., entering illegal values into data fields).
4. All test activities and results should be documented and distributed to the relevant IT and OT organizational units.

#### 5.3.1.8 Security

This section addresses security requirements, typically obtained from NIST, IEC standards, BPs, and other sources. We do not present all aspects but only emphasize certain general aspects of security requirements concerning SbD.

1. Identify as many security requirements as possible.

The organization can accomplish this by obtaining expert knowledge, gathering external information, and using techniques such as brainstorming, etc.

2. Translate non-functional requirements.

   Some of the requirements described in clause 1 are high-level security requirements (e.g., confidentiality, non-repudiation, integrity). Such requirements may be viewed as non-functional requirements; therefore, organizations might assume that they are trivial and neglect them. However, because these requirements are fundamental, they should be translated into functional requirements and included in the requirement list. For example, integrity, which is a non-functional requirement, is typically translated into access control, non-repudiation, physical protection, attack detection, etc.; each of these is further translated to a more technical level, e.g., access control can be defined by specific mechanisms such as biometrics, passwords, etc. And the last translation is to code.

3. Develop mutual understanding.

   Foster mutual understanding between security experts and the project team regarding security requirements. This effort should encourage a shared understanding of security requirements among all project participants.

4. Prioritize the security requirements.

   Once the list of security requirements is complete, the organization must prioritize them regarding what actions should be taken, the timeline, verification, testing methods, and so on. This information should be documented and stored in the KB (see Section 5.3.1.1).

5. Awareness

   In an organization, users often have varying levels of security awareness, which could affect their ability to adopt the SbD approach. As a result, it is essential to address users' lack of security awareness.

6. Interdependencies

   Typically, security requirements have interdependencies. For instance, implementing a firewall or IDS in the OT environment may impact the communication time between sensors and PLCs; encryption of data can compromise visibility, or safety and security. Therefore, it is essential to:

   a. Enhance understanding of the interdependencies between security requirements and their resolutions.
   b. Establish a testing methodology to assess those interdependencies and evaluate how they should be tested (see Section 5.3.1.7).

7. Changes

   Any change in a requirement (e.g., an application, infrastructure, or security requirement) necessitates a risk assessment and analysis to ensure that it is feasible to implement that change, along with checking for interdependencies and contradictions with other requirements.

8. Documentation and KB

   Some reasons why such KB is required include ensuring clarity and shared understanding, enabling change management and traceability, improving quality assurance and risk mitigation, and preserving history and organizational knowledge. Therefore, it is essential to:

   a. Document all security requirements, possible tests, and validation processes applicable to each requirement and the system. Documentation should include all of the building blocks relevant to each requirement.
   b. Develop a repository/KB (see Section 5.3.1.1) that contains all information related to security requirements and continuously update it during all development and maintenance phases.

#### 5.3.1.9 Other Subjects Related to the Requirement Phase

There are numerous standards and BPs that address topics such as remote access communication, segregation of duties, segmentation, perimeter security, backup and recovery, physical security, laws, and regulations. Therefore, we will not discuss them in this document.

## 5.4 Design Phase

Usually, the security requirements agreed upon by the organization serve as input to the design phase. As previously mentioned, we do not aim to change the lifecycle of SW, HW, or system development or the process of purchasing any of them. We do, however, raise some points concerning cybersecurity issues. We are confident that adopting this perspective will contribute to a higher level of cybersecurity and robustness against cyberattacks.

During the design phase, the designers/ architects:

a. Develop the technical specifications of the artifact, ensuring all details comply with the requirements.
b. Identify nonconformance or vulnerabilities that were not identified during the requirement phase.
c. When purchasing a product, determine how the product will be integrated into the organizational systems.
d. Determine how to mitigate security risks.

In the remainder of this section, we address some of the security issues that require emphasis. Other security issues are discussed in the literature (e.g., BPs, standards, academic and non-academic papers).

### 5.4.1 General Security Aspects Related to the Design Phase

This section outlines principles for embedding SbD during the design phase.

1. The organization's designers and security experts should analyze and discuss all aspects of security pertaining to the project.
2. The organization should ensure that its designers possess security awareness and receive relevant training.

3. If the organization's security experts lack the required knowledge and training, the organization should hire a third party that has the relevant security knowledge.
4. The CISO should approve the safeguard specifications and design before moving on to the next phase of the lifecycle.
5. Designers and security experts should analyze and evaluate potential misuse scenarios and malfunction situations that could lead to security vulnerabilities. Based on this analysis, they should develop appropriate countermeasures, which may require revising existing security requirements or introducing new ones.
6. A thorough design review is essential to identify potential flaws and ensure the system meets both functional and security requirements. We emphasize the following ideas about the design review.
    a. A process of review and verification should take place at the end of the design phase to uncover security flows, such as inadequate methods for raising alarms when malicious behavior is detected or a lack of encryption between two components.
    b. This process must be comprehensive and meticulously performed. At the end of the process, the designers, project managers, QA and security experts, and CISO should be confident that the design provides the highest level of security possible.
7. Decisions should be made regarding cryptographic protocols, standards, services, security frameworks, security mechanisms, etc.
8. Error handling mechanisms should be designed.
    a. Security mechanisms should be capable of functioning both during regular operation and when there are failures.
    b. To ensure the security of a system, security aspects should be considered in the design and implementation of all components of the system, including HW and SW. Therefore, designers and security experts need to:
        - Identify as many failure scenarios as possible during the design phase.
        - Ensure that there is full visibility in error or failure situations.
        - Establish a response mechanism for every potential failure scenario.
    c. In the event of failure, an error handling mechanism should take over.

d. The error handling mechanism is responsible for responding to and recovering from errors in SW applications, HW components, infrastructure, and communication systems.
   e. A well-designed ICS system (OT and IT) should be capable of predicting possible failures and errors, detecting and identifying failures, and initiating remediation activities.
9. All activities and artifacts from the design phase should be documented clearly and incorporated into the KB.

### 5.4.2 Other Security-Related Activities in the Design Phase

This section deals with security-related activities that were not discussed above.

#### 5.4.2.1 Enhanced Threat Modeling

Enhanced threat modeling enables us to identify various security threats, such as a malicious insider capable of changing a sensor's output.

1. After developing the basic threat model during the requirement phase, designers must increase their understanding of the security aspects pertaining to an artifact's design, including any previously uncovered security issues, possible vulnerabilities, and attack vectors.
2. In addition, designers should review the risk assessment and threat modeling (see Sections 4.4.11 and 5.3.1.1). This is key to ensuring that requirements' interdependencies are examined.

#### 5.4.2.2 Check and Recheck the Security Design Principles

Ensuring robust security begins with a careful examination of the design principles—and it doesn't end there; each principle must be revisited and validated throughout the development lifecycle. Complying with security requirements includes activities such as:

1. Ensuring that all security requirements are resolved during the design phase.

2. Verifying that each non-security requirement (e.g., application requirement, infrastructure requirement) does not violate the security architecture or create any vulnerabilities (see Section 4.4).
3. Determining the following:
    a. Whether the security controls are aligned with standards and BPs.
    b. Whether SW applications, HW, systems, etc. (developed and purchased) are aligned with standards and BPs.
    c. The ability to check and test the security measures developed or purchased. For example, ensure that there is an appropriate test to check whether saved passwords are encrypted, hashed, or saved in plain text.

## 5.5 Development Phase

This phase's purpose is to collect the outcomes from the previous phases (i.e., the requirement and design phases) and implement them as code.

With respect to purchased products, we assume some coding will be required to integrate them into the system (e.g., implementing visibility functions and integrating the product's log/historian into the ICS system).

With respect to in-house development, we only refer to the OT environment, i.e., levels 0 - 2 of the Purdue model, for the following reasons:

a. Those levels can be populated with IoT/ IIoT devices that run code.
b. Regardless of whether an analog sensor is used in level 0, it is connected to a device or component (e.g., PLC or IoT/IIoT device) that runs code.
c. Firmware is considered to be SW.

Therefore, the coding language in a PLC or firmware should still be viewed as a SW language.

Regarding the IT portion of an ICS, it is obvious that coding will need to be performed.

This section does not address secure coding techniques, as this topic has been covered in numerous BPs, books, and papers, as mentioned in previous sections.

As noted earlier, our objective is not to alter the existing SW, HW, or system development lifecycles, nor to modify their associated procurement processes. Instead, we draw attention to key cybersecurity considerations that must be addressed within those established frameworks. We contend that embracing this perspective will strengthen the overall security posture and improve resilience against potential cyberattacks.

### 5.5.1 Secure Coding

The organization should use a programming language that supports secure coding.

To the best of our knowledge, there are no BPs or standards for secure coding in the programming languages used in PLCs and firmware. Therefore, we encourage the energy sector to develop new programming languages that will have secure coding capabilities and support high availability. Alternatively, it may be necessary to use modern devices (e.g., IoT/ IIoT devices) that support programming languages and enable secure coding.

In this section, we do not aim to instruct coding experts on how to code. Rather, we present the following guidelines for integrating SbD and security features in coding practice:

1. Code should be developed in accordance with the relevant requirements and specifications (e.g., design specifications). Discrepancies should be resolved without neglecting SbD and cybersecurity requirements, mechanisms, and considerations.
2. Developers should pay particular attention to security aspects related to coding, such as secure coding and connections among modules.
3. When developing, developers should pay particular attention to testability, i.e., how the artifact (e.g., program) or purchased product is to be tested.
4. Comprehensive unit testing (i.e., the smallest piece of code that can be logically isolated in a system; see smartbear.com) should include non-security and security tests.
5. Version management and control tools should be used to transfer artifacts between one environment and another (e.g., from the development environment to the test environment). This applies to both OT and IT environments.

6. The transfer of code between environments should be approved by the project's management team, CISO, and the person responsible for the receiving environment (i.e., the test environment).

Note the following:

- We are aware that there are coding languages in the OT environment that do not have secure code language or mechanisms. When such coding languages are used, rigorous testing should be performed.
- HW components used in levels 0-3 of the Purdue model that will operate in an unprotected environment (e.g., outdoors) should be electromagnetically protected.

### 5.5.2 Code Review

Code review is considered a QA tool. However, the code review process can also be used to identify vulnerabilities and security bugs. Code reviews can be performed manually or automatically; these two methods are complementary and should be used together. Some of the benefits of code review, manual or automatic, are:

1. Find vulnerabilities (e.g., a part of the code that is not being used or a part of the code that, when used, will cause a vulnerability).
2. Identify incorrectly defined or used components (e.g., misused protocol).
3. Identify complex code that can be replaced with a simple alternative.
4. Identify design misinterpretations (e.g., the requirement phase requires 2FA, however, due to a glitch, users may be able to use 1FA sometimes).
5. Identify undesirable flows in the implementation logic and workflow.
6. Identify deficiencies and vulnerabilities in earlier phases (e.g., a function that was not assigned a security level in the design phase).
7. Analyze the flow of data among components.
8. Using automated tools saves time and effort; therefore, it is cost-effective.

Note the following:

- As part of its code verification process (manual and automated), the organization

Should use formal code verification tools.

- An automated code review tool should be able to analyze the code, both statically and dynamically.
- Static and dynamic code reviews should be performed in the development and testing phases.
- Dynamic code review should be incorporated into regular audits and security testing (see Sections 5.3.1.7 and 5.5.3) to identify unauthorized code modifications.

### 5.5.3 Testing

While it is not our aim to explain how test experts should perform their activities and what methodology they should use, in this section we highlight some aspects of testing that can enhance cybersecurity.

1. We assume that testing will be performed by experts, such as employees or third parties specializing in OT, IT, and communication system testing.
2. Applications, operating systems, HW, communication, and security testing experts should be familiar with the standards, BPs, and all relevant information pertaining to their area of expertise.
3. Testing experts should be provided with the organization's requirement documents (including design documents), the results of unit and integration tests, and code analysis results.
4. The testing process should be documented, including the test scenarios and results.
5. Testing information (e.g., test scenarios, method, and results) should be available to the project's designers, coding experts, test experts, and CISO.
6. The organization should set up a testing environment that encompasses and integrates its OT, IT, and communication environments. The testing environment should be as similar to the production environment as possible.
7. The testing environment should be isolated from the outside world, e.g., the Internet.

8. A dedicated test environment should be created for tests requiring a connection to the outside world (e.g., the Internet, business partners). This test environment must also be isolated from the organization's networks and protected similarly to the production environment.
9. The security tests to perform include the following:
   a. Penetration tests
   b. Robustness tests
   c. Fuzz tests
   d. Denial-of-service (DoS)/distributed denial-of-service (DDoS) attack tests
   e. Error handling tests
   f. Data protection tests
   g. Other testing scenarios in which an attempt to penetrate the system is made, with the aim of identifying the system's vulnerabilities.
10. Once all artifact discrepancies have been addressed and no vulnerabilities have been found for an extended period (e.g., a week) of thorough testing efforts, the testing will be deemed complete and certified by the project manager, CISO, and the individual responsible for testing.

### 5.5.4 Deployment in the Production Environment

1. The deployment of the artifact in the production environment consists of the following three steps.
   a. The SW, HW, and system are deployed in a pre-production environment. This will allow a series of cybersecurity tests (described in Section 5.5.3) to be performed on the artifact in an environment that is as close as possible to the production environment.
   b. The SW, HW, and system are deployed in the production environment.
   c. As soon as the artifacts have been deployed in the production environment, penetration testing, as well as other security tests, should be performed.

2. The project manager, CISO, and the person in charge of testing are responsible for declaring that the artifact can be transferred to the production environment.
3. All SW deployments should be performed through the version management and control system.

#### 5.5.5 General Recommendations Related to Testing

1. The energy sector ecosystem should establish an independent testing institute responsible for performing security testing and evaluating all COTS HW and SW used by organizations in the sector.
2. All test documentation and evaluation results should be made available to the ecosystem's members.
3. In addition to clause 1 above, every organization should conduct its own security tests on the same COTS HW and SW, as well as on its architecture and environment.

### 5.6 Maintenance

This section deals with maintenance activities (this also includes replacement activities). We only consider computerized components in the ICS environment (e.g., IoT devices, servers, SW applications, PLC, sensors containing CPU and memory). It is not our intent to cover all of the required maintenance activities. However, we describe a few activities that are considered important for supporting SbD. Nor do we dictate how an organization's maintenance experts should perform their activities and what methodologies should be used. However, we must emphasize that the organization should ensure that any maintenance activity does not endanger or harm the cybersecurity protection mechanisms in place.

At a minimum, the following cybersecurity activities should be performed:

1. Pre-processing activities

a. Ensure that the components undergoing maintenance have been tested according to organizational policies and procedures, e.g., in a testing environment or a near-production environment.
   b. Ensure that a fallback mechanism is available and has been tested in a near-production environment.
2. Post-processing activities
   a. Conduct a cybersecurity test in the production environment.
   b. Coordinate with the various organizational entities responsible for ensuring the success of the maintenance activity (e.g., validation of outcomes).

### 5.7 Disposal

International standards address IT component disposal, such as NIST SP 800-88 Revision 1 "Guidelines for Media Sanitization," 800-53 Revision 5 "Security and Privacy Controls for Information Systems and Organizations," and 800-171 Revision 2 "Protecting Controlled Unclassified Information in Nonfederal Systems and Organizations." Therefore, we do not elaborate further on this topic except to note that there is a requirement to impose those standards on all devices installed in levels 0-2 of the Purdue model.

## 6. Conclusions and Future Work

This document presents the requirements for implementing the SbD concept in ICSs, particularly in the energy sector. This work is driven by the assumption that the OT portion of these systems is on a path toward digitalization of its components, including sensors, actuators, and PLCs. Therefore, there is a need to both consider aspects of cybersecurity and implement them in the OT environment (with the necessary modifications), as they are implemented in the IT environment. As the transition to digitization begins, it is natural to incorporate cybersecurity via our proposed SbD framework.

As part of this framework, we strongly recommend establishing an ecosystem for all organizations involved in the energy sector. This ecosystem will have the authority and ability to enforce adherence to cybersecurity requirements and take a leadership role in ongoing efforts to improve cybersecurity across the energy sector.

This paper is structured into two main sections. The first focuses on non-technical requirements, including the role of senior management. The second covers the technical aspects, along with the SW development lifecycle.

Further research is necessary to establish comprehensive SbD methodologies that encompass the entire hardware lifecycle, integrate hardware and software within firmware, and address cybersecurity risks throughout the supply chain.

## Appendix A - List of Acronyms

| | |
|---|---|
| AES | Advanced Encryption Standard |
| BP | Best practice |
| CERT | Cyber emergency response team |
| CISO | Chief information security officer |
| COBIT | Control objectives for information and related technologies |
| COTS | Commercial off-the-shelf |
| CPU | Central processing unit |
| DES | Data encryption standard |
| DLP | Data leakage prevention |
| DMZ | Demilitarized zone |
| DNP3 | Distributed Network Protocol 3 |
| DDoS | Distributed denial-of-service |
| DoS | Denial-of-service |
| ECC | Elliptic curve cryptography |
| EPROM | Erasable programming read-only memory |
| HMI | Human-machine interface |
| HW | Hardware |
| ICS | Industrial control system |
| IDPS | Intrusion detection prevention system |
| IoT | Internet of things |
| IIoT | Industrial Internet of things |
| IT | Information technology |
| KB | Knowledge base |
| MFA | Multi-factor authentication |
| MTD | Moving target defense |
| NMS | Network management/monitoring system |
| OTP | One-time password |
| OT | Operational technology |
| PLC | Programmable logic control |
| QA | Quality assurance |
| RAM | Random access memory |
| RBAC | Role-based access control |
| ROM | Read-only memory |
| RSA | Rivest-Shamir-Adelman encryption method |
| SbD | Security by design |
| SDLC | Software development lifecycle |
| SIEM/SOC | Security incident and event management and security operations center |
| SIEM | Security incident and event management |
| SOC | Security operations center |
| SW | Software |
| TPM | Trusted platform module |
| UML | Unified modeling language |
| ZT | Zero trust |
| 1FA | One-factor authentication |
| 2FA | Two-factor authentication |